\documentclass[
aps,
amsmath,amssymb,
reprint,
superscriptaddress,
floatfix
]{revtex4-2}

\usepackage{dcolumn}%
\usepackage{bm}%
\usepackage[utf8]{inputenc}
\usepackage[T1]{fontenc}
\usepackage{mathptmx}
\usepackage{etoolbox}

\usepackage{graphicx}%
\graphicspath{./ }
\usepackage{amsmath,amssymb}
\usepackage{gensymb}
\usepackage[dvipsnames]{xcolor}

\usepackage[colorlinks=true,
citecolor=blue,
linkcolor=magenta]{hyperref}

\begin{document}

\preprint{AIP/123-QED}

\title{Single-cycle all-fiber frequency comb}
\author{Sida Xing}
 \email{sida.xing@colorado.edu}
 \affiliation{Time and Frequency Division, NIST, 325 Broadway, Boulder, Colorado 80305, USA}
 \affiliation{Department of Physics, University of Colorado, 2000 Colorado Ave., Boulder, Colorado 80309, USA}
\author{Daniel M.B. Lesko}%
 \affiliation{Time and Frequency Division, NIST, 325 Broadway, Boulder, Colorado 80305, USA}
 \affiliation{Department of Chemistry, University of Colorado, 215 UCB, Boulder, Colorado 80309, USA}
\author{Takeshi Umeki}%
 \altaffiliation[Permanent address: ]{NTT Device Technology Laboratories, NTT Corporation, 3-1 Morinosato Wakamiya, Atsugi, Kanagawa 243-0198, Japan}
 \affiliation{Time and Frequency Division, NIST, 325 Broadway, Boulder, Colorado 80305, USA}
\author{Alexander J. Lind}
\affiliation{Time and Frequency Division, NIST, 325 Broadway, Boulder, Colorado 80305, USA}
 \affiliation{Department of Physics, University of Colorado, 2000 Colorado Ave., Boulder, Colorado 80309, USA}
\author{Nazanin Hoghooghi}
 \affiliation{Precision Laser Diagnostics Laboratory, Department of Mechanical Engineering, University of Colorado Boulder, Boulder, CO 80309, USA}
\author{Tsung-Han Wu}
 \affiliation{Time and Frequency Division, NIST, 325 Broadway, Boulder, Colorado 80305, USA}
 \affiliation{Department of Physics, University of Colorado, 2000 Colorado Ave., Boulder, Colorado 80309, USA}
\author{Scott A. Diddams}
\email{scott.diddams@nist.gov}
 \affiliation{Time and Frequency Division, NIST, 325 Broadway, Boulder, Colorado 80305, USA}
 \affiliation{Department of Physics, University of Colorado, 2000 Colorado Ave., Boulder, Colorado 80309, USA}

\date{\today}%

\begin{abstract}
Single-cycle pulses with deterministic carrier-envelope phase enable the study and control of light-matter interactions at the sub-cycle timescale, as well as the efficient generation of low-noise multi-octave frequency combs. However, current single-cycle light sources are difficult to implement and operate, hindering their application and accessibility in a wider range of research. In this paper, we present a single-cycle 100 MHz frequency comb in a compact, turn-key, and reliable all-silica-fiber format. This is achieved by amplifying 2 µm seed pulses in heavily-doped Tm:fiber, followed by cascaded self-compression to yield 6.8 fs pulses with 215 kW peak power and 374 mW average power. The corresponding spectrum covers more than two octaves, from below 700 nm up to 3500 nm. Driven by this single-cycle pump, supercontinuum with 180 mW of integrated power and a smooth spectral amplitude between 2100 and 2700 nm is generated directly in silica fibers. To broaden applications, few-cycle pulses extending from 6 µm to beyond 22 µm with long-term stable carrier-envelope phase are created using intra-pulse difference frequency, and electro-optic sampling yields comb-tooth-resolved spectra. Our work demonstrates the first all-fiber configuration that generates single-cycle pulses, and provides a practical source to study nonlinear optics on the same timescale.
\end{abstract}

\maketitle

\section{Introduction}

\begin{figure*}[!ht]
    \centering
    \includegraphics[width=.9\linewidth]{ 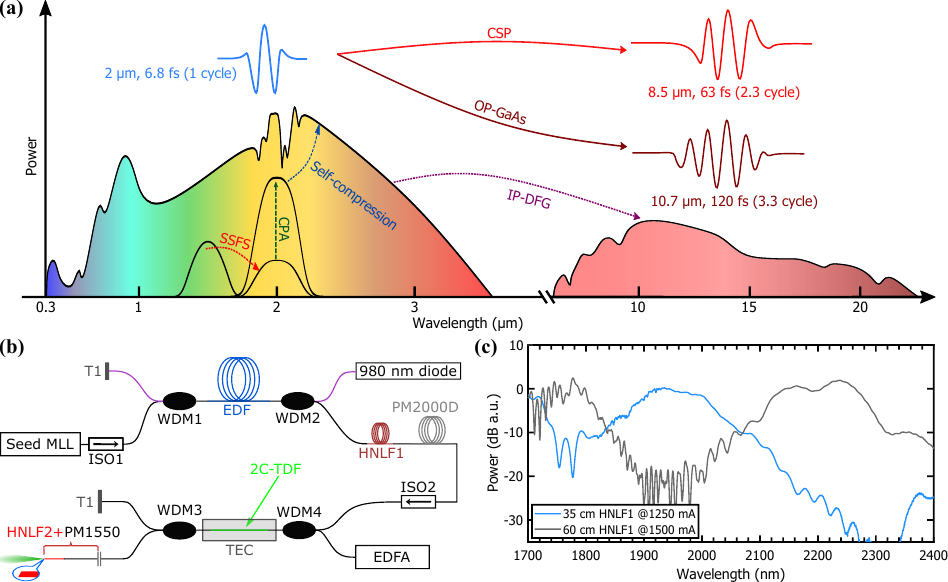}
	\caption{(a). Concept of the experiment creating MIR frequency combs based on a single-cycle all-fiber pump at 2 µm. (b). Experimental setup of the Tm-doped fiber amplifier. All optical components in this setup are polarization maintaining. T1: terminated port. MLL: mode-locked laser. ISO: isolator. WDM: wavelength division multiplexer. EDF: Er-doped fiber. HNLF: highly nonlinear fiber. 2C-TDF: double-cladding Tm-doped fiber. TEC: thermal-electric controller. (c). Self-frequency-shifted soliton in the 2 µm range which is used as a seed pulse (blue trace). A longer HNLF length and higher 980 nm diode pump power leads to a soliton shift up to 2.27 µm}
    \label{fig:laser-setup}
\end{figure*}
Since the invention of the first femtosecond lasers \cite{Shank1974}, continuous efforts have been invested into pushing the duration of laser pulses towards the limit of a single-cycle. Implementation of sub-two-cycle sources has been achieved using Ti:sapphire technology \cite{Rausch2008, Yamane2003}, filamentation \cite{Fuji2007}, and optical parametric amplification \cite{Chen2019}. In 2011, researchers demonstrated single-cycle pulse synthesis utilizing silica fibers \cite{Krauss2010}, and coherent pulse synthesis became another promising approach \cite{Manzoni2015}. Nevertheless, all-fiber single-cycle frequency combs with nJ pulse energy and high repetition rate ($>50$ MHz) are still absent. Applications exploring sub-cycle light-matter interactions, such as probing valence band electron motion \cite{Morimoto2021}, studying sub-cycle electron dynamics \cite{Kruger2011, Higuchi2017,Garg2020, Ludwig2020}, and sub-cycle current control \cite{Hanus2021} require nJ-level, near single-cycle pulses with precise carrier-envelope-phase (CEP) control/stabilization. In the more classical nonlinear optics field, the envelope-based propagation equations are expected to be able to simulate single-cycle pulses \cite{Brabec1997,Dudley2006,Conforti2010}, and possibly describe sub-cycle dynamics \cite{Genty2007}. However, experimental validation is hindered by the availability of single-cycle sources. Supercontinuum generation (SCG) from a single-cycle driving pulse remains to be an unexplored field both experimentally and theoretically. Intrinsically, single-cycle pulses have in-phase octave-spanning spectrum and high peak power ($>100$ kW). When generated at high repetition rate and sent into planar waveguides or optical fibers, frequency combs can be coherently and efficiently expanded to multi-octaves \cite{Corwin2003, Dudley2004}. In addition, such pulses can drive intra-pulse difference frequency generation (IP-DFG) for broadband mid-infrared (MIR) spectra. While the theory describing single-cycle pulse generation via self-compression in optical fibers has been developed in 2005 \cite{Foster2005}, the technology to make their availability widespread has not previously existed.

A fiber laser source of single-cycle frequency comb has the benefits of being turn-key, permanently aligned, robust and accessible for researchers outside the laser physics community. This could lead to new research areas and opportunities in multiple fields. Among all wavelengths, the Tm/Ho:fiber band (1900 nm to 2200 nm) is most promising and interesting. This wavelength resides at the edge of the MIR while affording good transparency with mature, reliable and commercially available silica fiber components \cite{Moulton2009, Seddon2010, Fermann2013, Rudy2014}, leading to unique advantages in terms of amplification, dispersion engineering, and access to still longer wavelengths. Over the past few years, 2 µm lasers reaching two optical cycles by self-compression were reported \cite{Gebhardt2017,Xing2020}, with pulses as short as 9.4 fs pulses being produced from an all-fiber frequency comb at 100 MHz repetition rate \cite{Xing2020}. Unfortunately, accumulated higher order dispersion (HOD, $\beta_3$ and higher orders) degrades the pulse temporal quality by inducing a pedestal and satellite pulses, making it difficult to reach single-cycle realm. 

In this paper, we present an all-fiber-integrated solution that realizes pulses at the limit of a single-cycle and 100 MHz repetition rate. The frequency comb spectrum covers greater than two octaves -- from 700 nm to 3500 nm. Figure \ref{fig:laser-setup}(a) illustrates the concept of this work. It is noteworthy that we achieve single-cycle pulses and this spectrum using cascaded self-compression exclusively in silica fiber, and without any additional optical elements or materials. This achievement is built on a novel amplifier with only 14 cm of core-pumped, double-clad Tm-doped fiber (2C-TDF) that provides approximately 460 mW of power. After chirped-pulse-amplification (CPA), cascaded self-compression in silica fibers leads to 6.8 fs pulses with 374 mW average power. The configuration uses only polarization-maintaining (PM) silica fiber components, and the output polarization extinction ratio (PER) is more than 15 dB. Full nonlinear simulations show excellent agreement with our experimental results, both in the spectral and temporal domains. Wavelengths greater than 2200 nm are typically considered beyond the conventional working range of silica fibers \cite{Tao2015}. However, we show that efficient supercontinuum (SCG) up to 3.5 µm is indeed feasible driven by single-cycle pulses, promising new opportunities in coherent SCG and filling the gap between experiment and simulations from eight years ago \cite{Laekgsgaard2013}. By simply propagating in an extra 2.5 centimeters of nonlinear fiber, we generate a smooth supercontinuum spectrum covering 2.1 µm to 2.6 µm, with more than 180 mW between 2.1 µm and 2.8 µm. Finally, by focusing the 6.8 fs pulse into a single crystalline CdSiP\textsubscript{2} (CSP) crystal, we generate 860 µW MIR spectrum centered at 8.5 µm through IP-DFG. The MIR pulse electrical field is retrieved by electro-optical sampling (EOS), yielding a pulse duration of about 63 fs, or 2.3 optical cycles. Similarly, pumping an orientation-patterned gallium asernide (OP-GaAs) crystal provides 270 µW MIR spectrum until 25 µm with a corresponding pulse duration of 120 fs (3.3 optical cycles). EOS averaging resolves the 100 MHz comb tooth structure.

\begin{figure*}[!ht]
    \centering
    \includegraphics[width=.9\linewidth]{ 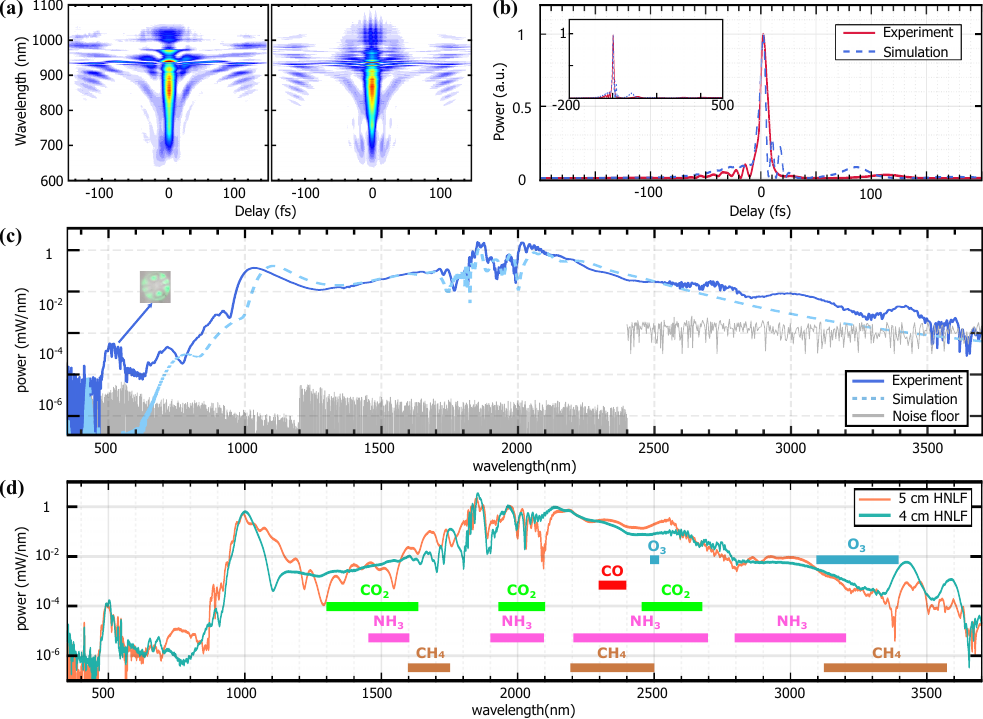}
	\caption{(a). Experimental (left) and reconstructed (right) SHG-FROG spectrogram. (b). Retrieved (solid) and simulated (dashed) pulse in temporal domain. Inset: Pulse temporal structure from -200 to 500 fs. (c). Experimentally recorded spectrum (solid) superimposed with simulation (dashed) of the 6.8 fs pulse. The noise floor of each spectrometer is plotted in gray. Note that the green light around 500 nm is in a higher order mode. (d). Output spectrum after further propagating the pulse in longer HNLF2 length.For spectroscopic applications, we also indicate the absorption range of several greenhouse gas and CO after 2 µm. In (c) and (d), all spectra are scaled to experimentally measured power.}
    \label{fig:HNLF-output}
\end{figure*}

\section{single-cycle All-fiber laser}
A straightforward fiber-integrated frequency shift from a 1.5 µm Er:fiber laser allows access to the Tm amplifier band near 2 µm \cite{Nishizawa2001,Kumkar2012,Klose2014}. As illustrated by Fig. \ref{fig:laser-setup}(a), the 2 µm acts as the seed to the 2C-TDF CPA. The most straightforward, general and broadband approach to suppress HOD accumulation is to reduce the CPA fiber length, which is a unique solution for Tm-doped fiber CPAs. Compared with Yb\textsuperscript{3+} and Er\textsuperscript{3+} ions, Al-doped silica glass can host much higher concentration of Tm\textsuperscript{3+} ions without clustering \cite{Sincore2018,Digonnet2001rare}, leading to commercially available highly-doped TDFs. Thus, we employ highly-doped Tm:fiber in the present CPA, successfully reducing the gain length by 14 times compared to our previous design \cite{Xing2020}. Short gain fiber length greatly suppresses HOD accumulation and outputs even higher average power, leading to better pulse quality, shorter pulse duration and higher peak power.

Figure.\ref{fig:laser-setup}(b) shows the complete laser layout, sitting on a 25 cm x 45 cm breadboard. A 1.56 µm, 100 MHz repetition rate Menlo Systems Figure-9 mode-locked laser (MLL) is amplified from 20 mW to 190 mW in a backward-pumped PM erbium-doped fiber (EDF). A dedicated port on the MLL is used to phase lock the two degrees of freedom of the frequency comb. After self-compression, the pulse duration decreases to about 35 fs when entering PM highly nonlinear fiber (HNLF1). At 1.55 µm, the HNLF1 group velocity dispersion (GVD) is -7.82 ps{\textsuperscript{2}}/km and nonlinear parameter is 10 (W-km){\textsuperscript{-1}}. The soliton self-frequency shift (SSFS) from HNLF1 seeds the thulium-doped fiber amplifier (TDFA). We selected the 1930 nm soliton wavelength for optimal CPA performance, shown by the blue trace of Fig. \ref{fig:laser-setup}(c). Notably, this seed soliton covers the bandwidth of TDF emission (1800 nm $\sim$ 2200 nm). When the HNLF1 length increases to 60 cm and diode power increases to 1 W (1500 mA), the soliton continues to shift up to 2.27 µm with 20 mW average power. This soliton could be directly used in NH{\textsubscript{3}}, CO and CH{\textsubscript{4}} spectroscopy. Further shifting is stopped by perturbation from high dispersion slope and/or higher loss at longer wavelength \cite{Lee2008}. The seed soliton is stretched with 1 m PM2000D fiber, reaching a net-normal dispersion in the TDFA and to avoid noticeable nonlinear effects in CPA. An isolator (ISO2) follows the PM2000D to minimize backward reflections and increase the seed PER. After the PM2000D, ISO2 and wavelength division multiplexer (WDM4), the 1930 nm seed soliton has about 20 mW average power with 93 nm 3 dB bandwidth. A small segment of 2C-TDF (PM-TDF-10P/130-HE, 10 µm core diameter) was spliced between WDM3 and WDM4. The 2C-TDF is pumped by a 1550 nm continuous-wave laser that is further amplified in an erbium-doped fiber amplifier (EDFA). At 1550 nm, the 2C-TDF has a small signal absorption of about 150 dB/m. Two splicing points with PM1550 introduce less than 0.8 dB loss, measured at 1310 nm. The 2C-TDF is buried inside thermopaste, and a solid-state thermoelectric cooler (TEC) keeps the temperature at 7 \degree C. This temperature is higher than the dew point for room temperature air at 30\% relative humidity. In the TDFA, a forward pumping scheme is used to red-shift the 2C-TDF gain center to 1930 nm, thanks to the stronger re-absorption and re-emission than backward pumping scheme. We optimize the 2C-TDF length to overlap the gain center with the seed pulse peak by cutback experiments. Cutback experiments were performed while the 2C-TDF is kept 7 \degree C, and the optimal length is 14 cm.  The average power directly after WDM3 is 457 mW measured by a thermal power meter. 

To reach a single-cycle, the output pulse requires more than one octave spectral bandwidth. This can be realized using a cascaded self-compression strategy \cite{Xing2020} after the output of WDM3. First, 58 cm PM1550 fiber balances the residual normal dispersion inside the TDFA. The length is determined by a cutback experiment and no spectral broadening is noticed when PM1550 length is less than 40 cm. The first compression stage results in approximately a 50 fs pulse duration, serving as the pump pulse for the second stage compression. In the second stage, an elliptical-core HNLF (OFS Optics) works as the self-compression fiber [HNLF2 in Fig. \ref{fig:laser-setup}(b)]. The total splicing loss between HNLF2 and PM1550, plus the Fresnel reflection loss, is about 1 dB. To avoid back reflection, the output of HNLF2 is polished to an 8$^{\circ}$ angle. Multiple cutback tests were performed on HNLF2 with 2 mm step size. We found the optimal length to be 2.6 cm - which is the same as our simulation prediction. The total output power after HNLF2 power is 374 mW for a pump power of 2.34 W, leading to 17\% efficiency. As all components in Fig. \ref{fig:laser-setup}(b) are PM, more than 15 dB PER is measured at the output of HNLF2 after a broadband polarizer. TDF has a higher absorption cross section at 1.6 µm \cite{Digonnet2001rare}, which could be used to shorten the TDF and increase efficiency. However, the current limiting factor of the total fiber length comes from the PM fiber splicer, not the pump wavelength.

An off-axis parabolic (OAP) metal mirror (0.2 dB loss) collimates the laser output. The pulse duration is characterized by a second-harmonic-generation frequency resolved optical gating (SHG-FROG) setup that uses all reflective optics, then further confirmed by simulation using the same HNLF2 parameters as shown in \cite{Xing2020}. After collimation, the pulse front tilt can be perfectly compensated with a 1 mm fused silica window. We show the experimentally recorded and numerically reconstructed FROG spectrogram in Fig. \ref{fig:HNLF-output}(a) over a $\pm$150 fs range, indicating a good agreement. From the spectrogram, the entire central/main pulse is in phase. To estimate the pulse quality, the FROG scan is increased to a $\pm$800 fs range. Figure \ref{fig:HNLF-output}(b) shows the FROG retrieved and simulated pulse in the temporal domain with 1\% reconstruction error, defined by equation (A.2) in \cite{Rhodes2013}. Negligible energy exists beyond -200 fs or +500fs, so we only plot -200 fs to 500 fs [the inset of Fig. \ref{fig:HNLF-output}(b)].  Zooming into $\pm$200 fs, the experimental pulse duration is 6.8 fs or 1.05 optical cycle at 1930 nm. The experimental pulse matches well with simulated pulse (6.7 fs) in both pulse shape and duration. This TDFA generates remarkably minor satellite pulses and concentrates more power in the main pulse. Notably, the pulse has very little pedestal, owing to the small HOD accumulation. More than 50 \% of power is in the central pulse, leading to higher than 215 kW peak power. The ratio of the integrated relative intensity noise, $\sigma$, is 0.03\% (100 Hz to 10 MHz), detailed in the supplementary material (S1). 

To record the compressed output spectrum, we used three spectrometers for visible, near-infrared and MIR band. Yokogawa 6374 and 6375 cover the spectrum from 350 nm to 1200 nm and 1200 nm to 2400 nm, respectively. An ArcOptix MIR spectrometer together with 2750$\pm$250 nm and 2.8 µm to 5.3 µm bandpass (BP) filters covers the remaining range. The collimated laser beam passes the 2750 nm BP filter at about 45\degree incident angle, shifting the BP filter center to about 2500 nm. In Fig. \ref{fig:HNLF-output}(c), we show the recorded spectrum at 374 mW output power, together with noise floors of the spectrometers. No residual 1.55 µm pump exists at any output levels. Self-compressed pulses typically yield a structured central region of the spectrum and smooth wings favoring the long wavelength side \cite{Foster2005, Amorim2009, Gebhardt2017, Gaida2018, Butler2019a, Xing2020}. Such distinct features still exist in both experiment and simulations when self-compression yields single-cycle pulses [Fig. \ref{fig:HNLF-output}(c)], also predicted by other group \cite{Foster2005}. These "fingerprints" can serve in empirical estimation of optimal HNLF length for self-compression. The recorded spectrum covers more than two octaves directly from this all-fiber system. The carrier-envelope-offset frequency has more than 30 dB signal-to-noise ratio at 100 KHz resolution, detailed in Supplemental material S3. Surprisingly, the spectrum expands even beyond 2.4 µm and continues up to 3.5 µm. Considering the high peak power, short propagation length, silica glass transmission up to 4 µm \cite{Tao2015} and removing of OH bonds in optical fibers \cite{Horiguchi1976,Thomas2000}, it is reasonable to observe the spectrum beyond 3 µm. As a result, one can expect that a longer HNLF2 length might generate even more power after 2.4 µm. The peaks at 2.7 µm, 3 µm and 3.5 µm are not predicted by our modeling - they are most likely originate from dispersive waves \cite{Laekgsgaard2013} or four-wave-mixing. 

We simulated the single-cycle pulse formation using the generalized nonlinear Schrodinger equation (GNLSE) detailed in \cite{Dudley2006}, including the Raman response, shock term and up to 6\textsuperscript{th} order of dispersion. Our past study \cite{Xing2020} showed that the HNLF2 GVD is -13 ps\textsuperscript{2}/km and nonlinear parameter is 3.1 (W-km)\textsuperscript{-1}. Using the GNLSE, our simulation covers the 2 µm seed soliton after HNLF1 until the output of HNLF2. In Fig. \ref{fig:HNLF-output}(c), the simulation results are superimposed with the experimental spectrum. The simulation shows good agreement with the experiment, even at single-cycle pulse duration and reproduces all features at short wavelength side until 400 nm. This implies that the nonlinear envelope equation is still valid at a single-cycle, and that the GNLSE approach is precise to use as a design tool to model pulse formation in the single-cycle domain. It should be noted that the green light is produced in a higher order mode and cannot be predicted by our model. In addition, our power sweep from 200 mW up to 374 mW (8 cycles to single-cycle), shows good matching between numerical (GNLSE simulation) and experimental (FROG retrieval) results, detailed in Supplementary material (S2). Nevertheless, a more precise simulation of single-cycle pulse formation and further propagation will greatly benefit from complete knowledge of HNLF properties. Based on previous researches \cite{Brabec1997,Karasawa2001}, the important factors are: fiber geometry, the W-shaped dispersion profile \cite{Yamamoto2016}, wavelength-dependent effective area, wavelength-dependent propagation loss and dispersion fluctuations \cite{Kuwaki1990,Farahmand2004}. Those parameters might help to correctly identify the origin of the three peaks after 2.4 $\mu$m [Fig. \ref{fig:HNLF-output}(c)]. 

It has been widely accepted that significant frequency comb coverage beyond 2.2 µm is not feasible in an all-silica fiber system \cite{Petersen2014, Sorokina2014, Baumann2019}. This is because: 1) coherent SCG is hindered by the increased propagation loss in silica fibers; 2) rare earth emission beyond 2.2 µm is strongly suppressed by multi-phonon relaxation \cite{Digonnet2001rare}. The greenhouse gases and atmospheric pollutants have absorption peaks between 2 µm to 2.6 µm, making this an important spectroscopic region \cite{Bovensmann1999, Ehret2008, Canty2015}. To extend fiber laser spectrum beyond the silica window, people have invested significant effort in soft-glass optical fibers \cite{Tarnowski2016, xing2018linearly, Yao2018} and planar waveguides \cite{Singh2018,Carlson2019,Guo2020}. When pumped at 1.5 µm, Si\textsubscript{4}N\textsubscript{3} (SiN) is a promising waveguide platform for this purpose, and dual-comb spectroscopy at 2 µm to 3 µm has been demonstrated \cite{Baumann2019,Guo2020}. Still, an all-fiber source providing flat and bright spectrum in this range can be a more integrated, accessible, and robust solution for frequency comb spectroscopy. In addition, the intrinsic high mode quality of a single mode fiber allows easy SCG collimation with a commercial OAP collimator. With our single-cycle source, we show efficient SCG generation beyond 2.2 µm in a small segment of silica fiber. Both the short pump pulse and short nonlinear fiber lead to better SCG coherence \cite{Dudley2004}, which is critical for frequency comb spectroscopy. 

\begin{figure}[ht!]
       \includegraphics[width=\linewidth]{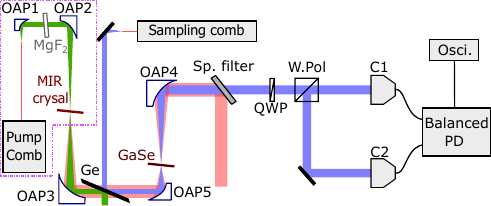}
        \centering
        \caption{Experimental setup for MIR frequency comb generation and EOS pulse characterization. The boxed part generates the MIR power and the rest of setup is for EOS characterization. OAP: off-axis parabolic mirror; MIR crystal: OP-GaAs or CSP crystal; Sp. filer: short pass filter; QWP: quarter-wave plate. W.Pol: Wollaston polarizer; C1\&C2: collimator 1 and 2; Balanced BP: balanced photodetector; Osci: oscilloscope.}
        \label{fig:EOS-setup}
\end{figure}
By simply increasing the length of HNLF2 length, we found the optimal length for flat SCG spectrum is 5 cm. In Fig. \ref{fig:HNLF-output}(d), we show the power spectral density of the SCG from two HNLF segments at the same EDFA pump power with 1 nm resolution. The average output power is 340 mW from 5 cm of HNLF and 350 mW from 4 cm of HNLF, which is a result of photon energy loss due to Raman interaction. The 5 cm HNLF2 SCG has two flat regions, covering 2 µm to 2.6 µm and 2.8 µm to 3.2 µm.  A 5 mm ultraviolet fused silica window placed in the beam path can block the spectrum beyond 2.8 µm \cite{Tao2015}, proving the spectrum is not an artifact. The total power from 2.1 µm to 2.8 µm is 180 mW, and the region after 2.8 µm contains 200 µW power. This SCG directly from HNLF provides higher power than SiN waveguides in the range of 2 µm to 2.8 µm and comparable power beyond 2.8 µm \cite{Guo2020}. Tapering of the HNLF could be a method to improve the dispersive wave efficiency at longer wavelengths. 

\begin{figure*}[!ht]
    \centering
    \includegraphics[width=.9\linewidth]{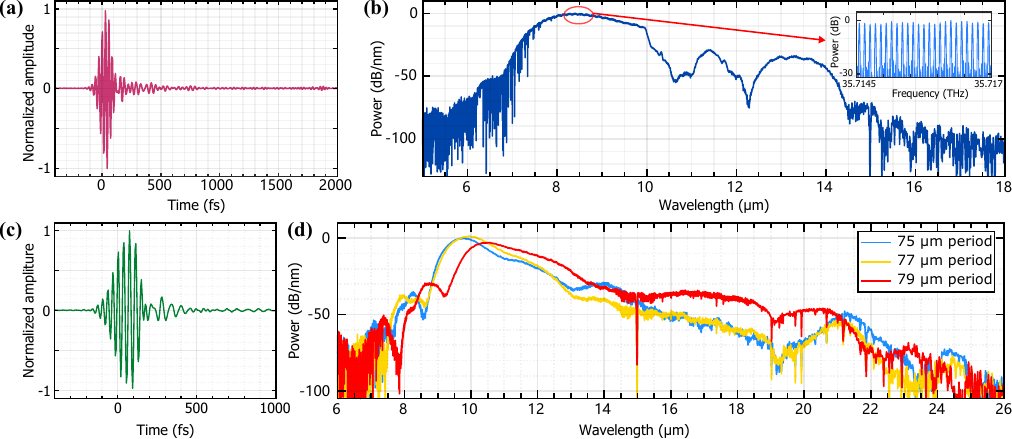}
    \caption{(a). The electrical field of the MIR pulse from CSP. The MIR pulse is about two optical cycles. (b). The MIR pulse spectrum from CSP after Fourier transform. The resolution is 1 GHz. (c). The EOS retrieved electrical field from OP-GaAs. This sample has a period of 79 µm. (d). MIR spectrum from OP-GaAs with three poling periods. The spectra come from Fourier transform of corresponding electrical field with 1 GHz resolution. CO\textsubscript{2} absorption line at 15 µm is clear. Inset: resolved comb lines spaced by 100 MHz near the peak of the MIR spectrum.}
    \label{fig:MIR-EOS}
\end{figure*}

\section{MIR generation and characterization}

In this section, we show MIR IP-DFG using two types of nonlinear crystals, OP-GaAs and CSP, as an immediate application demonstration of the single-cycle pump. The comb teeth in the MIR pulses are resolved using EOS. Therefore, this also serves as direct proof that the pump is a comb source with long-term stable repetition rate and carrier-envelope-offeset frequency. 

Coherent light sources in the MIR frequency range (3 µm to 25 µm) are critical for molecular spectroscopy as applied to fundamental materials science and chemistry \cite{Bjork2016,Fleisher2014}, environmental monitoring \cite{giorgetta2020open,Herman2021}, the determination of protein structure \cite{Klocke2018}, combustion analysis \cite{Makowiecki2020, Abbas2019} and medical diagnostics such as breath analysis \cite{Henderson2018}. With high peak power and a broad spectrum, the output pulse is suitable to produce MIR pulse through IP-DFG, which we explore with both OP-GaAs and CSP. Other crystals, like GaSe and ZnGeP\textsubscript{2} crystals were studied by several groups \cite{Gaida2018, Vasilyev2019a, Lesko2021} and an overview of different crystals can be found in \cite{vodopyanov2020book}. The boxed part of Fig. \ref{fig:EOS-setup} illustrates the experimental setup for IP-DFG. After OAP1, the collimated beam passes through 5 mm of MgF\textsubscript{2} to provide extra dispersion tunability for higher IP-DFG efficiency. Two OAP focal lengths are used to maximize the MIR generation in different crystals. An OAP with 2.54 cm focal length focuses the laser to 9 µm diameter at full-width-half-maximum (FWHM), while a focal length of 3.8 cm provides a FWHM of 12 µm. 

CSP has good MIR transmission, high damage threshold, good thermal conductivity, large bandgap, reduced loss for 2 $\mu$m pumps and high nonlinearity \cite{Petrov2012,Schunemann2016}. In addition, its relatively flat refractive index in the MIR window is beneficial for broadband MIR spectrum generation \cite{Schunemann2016}. A 500 µm thick type-II CSP crystal is used in this setup with 9 µm focal spot size. A three dimensional translation stage, together with a rotation mount, gives us full spatial tunability of the CSP crystal. OP-GaAs, on the other hand, allows MIR transmission beyond 15 µm and enables quasi-phase matching \cite{Schunemann2016,Boyko2018}, and we also investigated several 1 mm-thick OP-GaAs crystals of different period. In this case, the 2 µm pump avoids the two-photon-absorption process in OP-GaAs. OAP3 collects the output MIR and residual pump light. The OP-GaAs crystal is anti-reflection coated for 6$\sim$15 µm. Notably, this is the first experiment studying IP-DFG in OP-GaAs crystals. After a 4.5 µm longpass filter, we used a thermal power meter to measure the MIR power and analyze the spectra with dual-comb EOS \cite{Kowligy2019}. The same MIR spectrometer mentioned previously is used for real-time optimization of MIR power.

Figure \ref{fig:MIR-EOS}(a) shows the MIR pulse electrical field after approximately 30 min averaging (or 2\textsuperscript{14} waveforms) of the dual-comb EOS, indicating a near transform-limited electric field, as detailed in supplementary material S4. The pulse duration is about 63 fs, or close to two optical cycles at 8.5 µm. The corresponding spectrum is shown in Fig. \ref{fig:MIR-EOS}(b). The MIR power is measured to be 860 µW after the longpass filter without correction of Fresnel reflection. At the -20 dB level, the spectrum covers from 7 µm to 10 µm. The dips at about 10.5 µm and 12 µm are due to impurity contamination in CSP \cite{Schunemann2016}. Water absorption lines are recorded around 6 µm to 7 µm. This spectrum covers the absorption bands of many molecules such as O\textsubscript{3}, CH\textsubscript{3}OH and OCS, and the dip at 15 µm is due to CO\textsubscript{2}. The inset of Fig. \ref{fig:MIR-EOS}(b) shows a zoomed comb structure at the peak of the MIR spectrum with 100 MHz line-spacing after averaging for five hours ($2^{16}$ samples).  

The transmission window of CSP goes up to 10 $\mu$m, as indicated in \ref{fig:MIR-EOS}(b). In order to generate MIR combs at longer wavelength, we switched to OP-GaAs crystals. The OP-GaAs crystals lead to most efficient MIR IP-DFG with 5 cm of HNLF2, whose spectrum is shown in Fig. \ref{fig:HNLF-output}(d). At 5 cm HNLF length, we estimated the peak power to be more than 110 kW. Meanwhile, this pump pulse provides more power at the phase-matched signal wavelength \cite{Vodopyanov2005,Skauli2003}, leading to more efficient conversion than directly sending the 6.8 fs pump. Three-photon absorption \cite{Heckl2016} is not observed, as 3.8 cm focal length OAP gives higher MIR power than OAP with 5 cm focal length. In addition, MIR power is stable over at least 4 hours, so we believe no photodarkening occurred in the crystal. In Fig. \ref{fig:MIR-EOS}(c) and (d), we show the electric field and spectrum from OP-GaAs. The MIR pulse duration is about 120 fs for all three samples. In Fig. \ref{fig:MIR-EOS}(d), we show the  MIR spectrum calculated from the Fourier transformed EOS signal. These OP-GaAs samples are coated with anti-reflection (AR) coating for 2 µm band and the MIR band up to about 14 µm. However, the AR coating becomes less effective beyond this wavelength, giving rise to the Etalon fringes seen in the spectra of [Fig. \ref{fig:MIR-EOS}(d)]. The vibration of CO\textsubscript{2} at 15 µm is well-pronounced. All three samples generate about 270 µW MIR power measured after the longpass filter. The spectrum covering 9 µm to 11 µm is highly interesting for frequency comb spectroscopy, especially for gases with fundamental resonant frequency in this range, like CH\textsubscript{3}OH, NH\textsubscript{3} and CH\textsubscript{3}Br.
\section{Further work and conclusion}

To conclude, we presented an all-PM-fiber approach to provide 2 µm single-cycle optical pulses using only commercial components. To the best of our knowledge, this is the first time that single-cycle pulses are created directly at 100 MHz repetition rate using an all-fiber configuration. This single-cycle source can be CEP-stabilized \cite{Okubo2018} and accessible to the most researchers in the optics community with turnkey and maintenance-free operation. Due to the high doping concentration of 2C-TDF, we managed to use the most straightforward, most broadband and most reliable approach to suppress the HOD accumulation. This configuration is general, i.e. it is possible to scale up the peak power, repetition rate and even the operation wavelength. For example, commercially available TDFs can have small signal absorption of 300 dB/m at 1.55 µm, so it is possible to scale the average power in a similar TDFA length as would be required to achieve comparable pulse energy at GHz repetition rates. Similarly, using 1.6 µm pump and/or TDF with higher doping concentration can provide much higher average power. Such improvements could explore the possibility of fiber-based sub-cycle pulse generation using self-compression. Customized fibers, such as photonic crystal fiber or fluoride fiber, for the self-compression stage should make MW peak powers accessible in an all-fiber configuration. We would also like to point out that Tm-doped or Ho-doped silicate and germanate fiber can reach higher doping concentration \cite{Lee2015, Kuan2016}. Splicing between silica and soft-glass fibers has already been demonstrated \cite{Wang2009, Thapa2015}. As a result, a similar approach could apply to these gain fibers as well, leading to sub-2 cycle pulses at 2.1 µm range. 

We then demonstrated two immediate applications of the single-cycle driving pulse - SCG and IP-DFG. Pumped by the single-cycle pulses, we further introduce an all-fiber approach to efficiently provide frequency comb spectra up to 3.5 µm, with the possibility to improve the efficiency. Notably, the spectrum almost covers the entire MIR transmission window of silica glass. Coherent SCG in silica fibers with such coverage has only been showed numerically \cite{Laekgsgaard2013}, proving the possibility of new research aspects with single-cycle driving pulses. With 180 mW and a flat spectrum in 2.1 µm to 2.8 µm, this all-fiber system should be ideal for frequency comb spectroscopy aimed at monitoring greenhouse gases and atmospheric pollutants in this spectral window. The power spectral density of the comb exceeds, or is on par with some of the most carefully designed waveguides for this wavelength window. Special fibers, such as silica photonic crystal fibers or soft glass fibers, can be spliced with low loss to our laser and lead to broader spectrum with higher power beyond 3 µm. In addition, we anticipate more power to be generated beyond 2.8 µm by tapering the silica PM-HNLF. 

Finally, we demonstrated MIR frequency comb generation using IP-DFG. With CSP crystal, we generated 860 µW, 63 fs CEP stable pulse centered at 8.5 µm. OP-GaAs, with better transmission in MIR, yields spectrum centered from 9.5 µm to 11 µm, with 270 µW power. Using EOS, we can easily observe the CO\textsubscript{2} resonance in only 3 m beam path length. The MIR spectrum centered around 10 µm is particularly interesting for high sensitivity spectroscopy. Besides, the resolved MIR comb lines also showed the stabilized commercial frequency comb can be converted to single-cycle pulses in all-fiber configuration with low noise and timing jitter. The single cycle fiber laser and the IP-DFG setup is operated for at least five hours per day since September of 2019. The laser output power, spectrum, pulse duration and PER remain the same and no maintenance has been performed. The IP-DFG spectrum and power remain the same for both OP-GaAs and CSP crystals.

In pushing the limits of pulse duration and spectral bandwidth, our effort has revealed the need for additional future research in several areas. First, improved simulation tools will be essential for accurate simulation of the single-cycle pulse driven SCG or IP-DFG. For example, for MIR generation, more work is needed to further improve the efficiency from OP-GaAs and understand the pulse propagation dynamics in the crystal. We also note that our simulations indicate the possibility of achieving sub-cycle pulses by self-compression. However, more work is required both theoretically and experimentally to  characterize and understand the associated nonlinear optics in the sub-cycle regime.  Second, when using our laser with free space components, pulse front tilt is inevitable due to the two-octave spectrum. At 6.8 fs pulse duration, it is important to avoid or compensate the pulse front tilt. And finally, it is well-known that EDF can induce extra dispersion around its gain/absorption peak \cite{Romagnoli1992,Hickernell1993}. Currently, no resonant dispersion from TDF was noticed - it is absent in both interferometry dispersion measurement \cite{Kharitonov2016} and MLL with heavily-doped TDF (300 dB/m pump absorption) \cite{Qiao2019}. However, our configuration is different, having $20\times$ gain within 14 cm gain fiber. The contribution of TDF resonant dispersion in our system remains to be explored. %

\section*{Supplementary material}
See the supplementary material for supporting information.

\section*{Acknowledge}
The mention of specific companies, products, or trade names does not constitute an endorsement by NIST. The authors thank Ian Coddington and David Carlson for their comments; Esther Baumann for providing the highly nonlinear fiber (HNLF2); John Dudley for the helpful discussions about simulation; Peter Schunemann and Kevin Zawilski for providing the CSP crystal. The project is funded by the Air Force Office of Scientific Research (FA9550-16-1-0016); NIST Physical Measurement Laboratory; Defense Advanced Research Projects Agency (SCOUT).
\\
\\The authors declare no conflicts of interest.
\\
\section*{Data Availability Statement}
The data and code that support the plots within this paper and other findings of this study are available from the corresponding authors upon reasonable request.
\newpage
\bibliography{2umTDF.bib}

\end{document}


\preprint{AIP/123-QED}

\title{Supplementary material to: \\Single-cycle all-fiber frequency comb}
\author{Sida Xing}
 \email{sida.xing@colorado.edu}
 \affiliation{Time and Frequency Division, NIST, 325 Broadway, Boulder, Colorado 80305, USA}
 \affiliation{Department of Physics, University of Colorado, 2000 Colorado Ave., Boulder, Colorado 80309, USA}
\author{Daniel M.B. Lesko}%
 \affiliation{Time and Frequency Division, NIST, 325 Broadway, Boulder, Colorado 80305, USA}
 \affiliation{Department of Chemistry, University of Colorado, 215 UCB, Boulder, Colorado 80309, USA}
\author{Takeshi Umeki}%
 \altaffiliation[Permanent address: ]{NTT Device Technology Laboratories, NTT Corporation, 3-1 Morinosato Wakamiya, Atsugi, Kanagawa 243-0198, Japan}
 \affiliation{Time and Frequency Division, NIST, 325 Broadway, Boulder, Colorado 80305, USA}
\author{Alexander J. Lind}
\affiliation{Time and Frequency Division, NIST, 325 Broadway, Boulder, Colorado 80305, USA}
 \affiliation{Department of Physics, University of Colorado, 2000 Colorado Ave., Boulder, Colorado 80309, USA}
\author{Nazanin Hoghooghi}
 \affiliation{Precision Laser Diagnostics Laboratory, Department of Mechanical Engineering, University of Colorado Boulder, Boulder, CO 80309, USA}
\author{Tsung-Han Wu}
 \affiliation{Time and Frequency Division, NIST, 325 Broadway, Boulder, Colorado 80305, USA}
 \affiliation{Department of Physics, University of Colorado, 2000 Colorado Ave., Boulder, Colorado 80309, USA}
\author{Scott A. Diddams}
\email{scott.diddams@nist.gov}
 \affiliation{Time and Frequency Division, NIST, 325 Broadway, Boulder, Colorado 80305, USA}
 \affiliation{Department of Physics, University of Colorado, 2000 Colorado Ave., Boulder, Colorado 80309, USA}

\maketitle

\section{Relative intensity noise}

\begin{figure}[!ht]
    \renewcommand{\thefigure}{S\arabic{figure}} 
    \centering
    \includegraphics[width = 13cm]{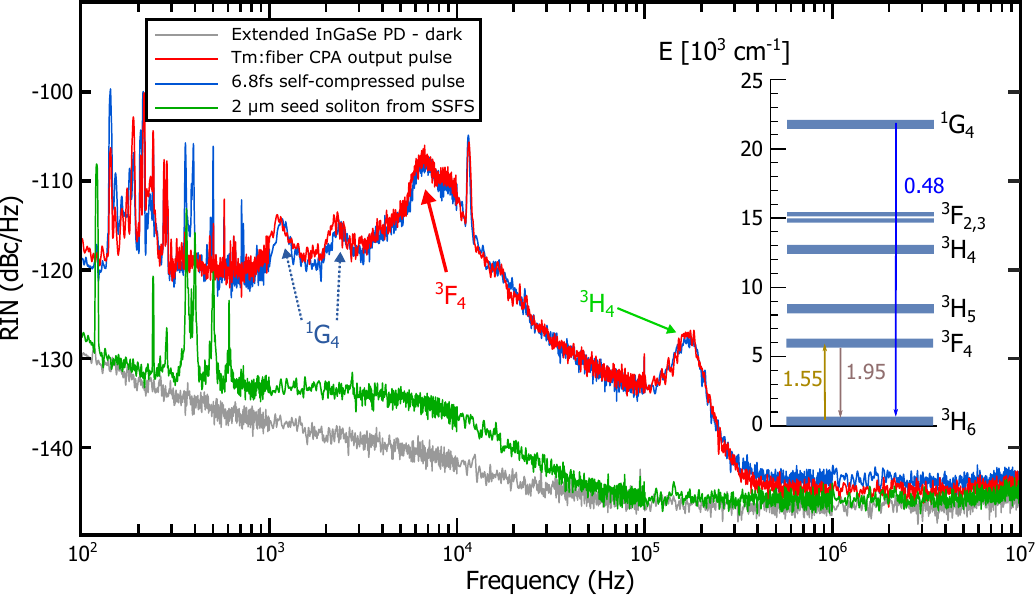}
	\caption{RIN measurement at different locations of the laser setup. We also associate the RIN maxima with the most probable excited states; Inset: simplified Tm\textsuperscript{3+} energy levels in silica host. The pump/emission transition and blue fluorescence transitions are highlight. The wavelength units are $\mu m$ in the transition.}
    \label{fig:rin}
\end{figure}

Our amplifier design provides very high gain per unit length and utilizes heavily doped Tm:fiber. Thus, it is necessary to perform a relative intensity noise (RIN) characterization of the single optical cycle pulse. In order to determine the origin of RIN, we perform the RIN measurement at different locations along the amplifier. The RIN measurement is performed with an extended InGaAs amplified detector, covering 900 nm to 2600 nm. Figure \ref{fig:rin} shows the recorded RIN from 100 Hz to 10 MHz. The InGaAs dark noise with no light illuminating it is recorded as a reference, shown by the blue trace. This provides the detector noise floor for RIN measurement. The red trace shows the RIN of the seed pulse, generated from soliton self-frequency shift. Its noise ratio $\sigma$ is only 0.002\% integrated from 100 Hz to 10 MHz. This low RIN value is owing to the short driving pulse (35 fs) and highly nonlinear fiber length (35 cm). The orange and green traces are the RIN after the Tm:fiber amplifier and after the self-compression to 6.8 fs, respectively. Both show nearly identical RIN, and have $\sigma = 0.03\%$ from 100 Hz to 10 MHz. The RIN from the Tm:fiber chirped pulse amplifier (CPA) contributes almost the entire RIN in the single cycle pulse. 

It is, therefore, important to identify the source of RIN from the Tm:fiber. The energy levels of Tm-doped silica fibers are plotted in the inset. During the CPA operation, we noticed clear blue/purple fluorescence from Tm:fiber, corresponding to the transition $^1G_4\rightarrow$ $^3H_6$ \cite{Peterka2011,Sincore2018}. All radiative levels below $^1G_4$ can be excited and contribute to the RIN. The most pronounced peak is associated with the transition $^3F_4\rightarrow$ $^3H_6$. The lifetime of $^3F_4$ level decreases with higher Tm doping concentration \cite{Sincore2018} and in our case it is about 200 $\mu s$. The maxima at about 10 kHz comes from $^3H_4\rightarrow$ $^3H_6$ \cite{Peterka2011}. Either or both maximas around 1 kHz correspond to the $^1G_4$ lifetime, but the radiative transition may not return directly to the $^3H_6$ level. Two simple approaches can help to suppress the RIN, 1) set the solid-state themo-electric controller to lower temperature; 2) pump the Tm:fiber with 1630 nm light. 
\newpage

\section{Pulse evolution with output power}

\begin{figure}[!ht]
    \renewcommand{\thefigure}{S\arabic{figure}} 
    \centering
    \includegraphics[width = 13cm]{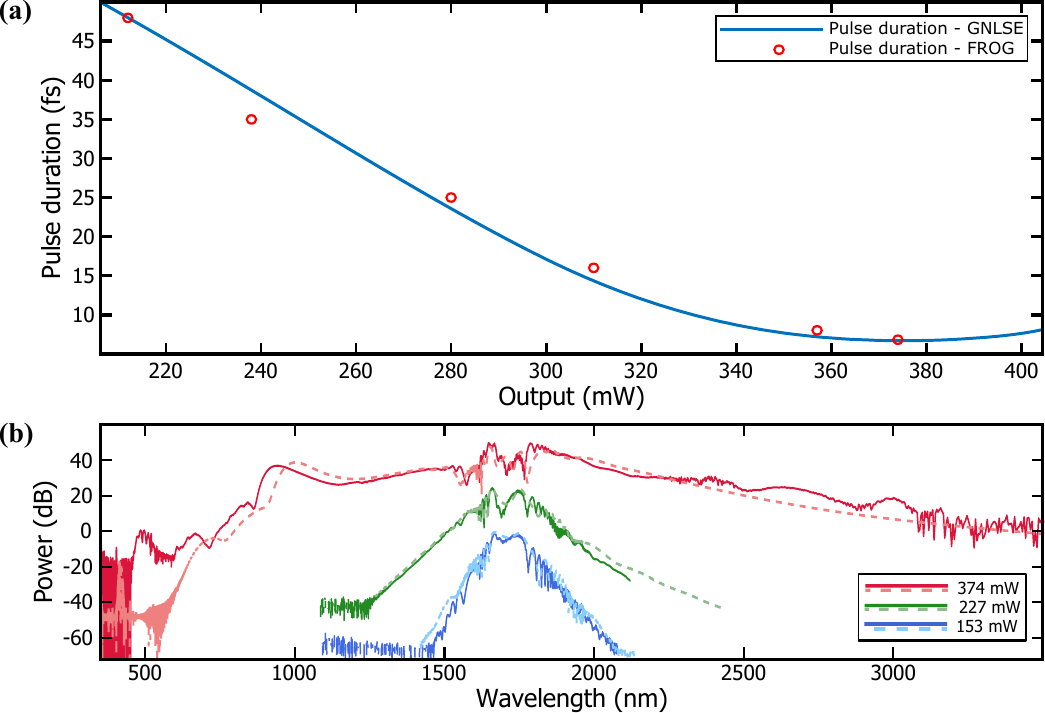}
	\caption{(a) Experimentally retrieved pulse duration from FROG (red) and simulated pulse duration (FROG) from GNLSE as a function of output power. In this experiment, our maximum output power is 374 mW.(b) Simulated and experimental spectra at various output power levels. For better illustration, each curve is shifted by 20 dB. }
    \label{fig:duration}
\end{figure}

In this section, we show a power scan and the corresponding pulse duration from 220 mW to 374 mW with results from both simulation and experimental retrieval. This is achieved by varying the pump power to the Tm:fiber amplifier; e.g. varying the CW erbium-doped fiber amplifier (EDFA) power. Figure \ref{fig:duration}(a) shows the pulse duration evolution from 50 fs ($\sim$8 cycles) down to 6.8 fs ($\sim$1 cycles). The experimental points starts from 31 dBm to 33 dBm EDFA pump power, with 0.5 dB step size. The last two data points correspond to EDFA power at 33.3 dBm and 33.5 dBm. The simulated pulse duration from GNLSE matches the retrieved pulse duration at all power levels and no abrupt change exists in neither method, in accordance with previous reports \cite{Baltuska1999,Brabec1997}. In Fig. \ref{fig:duration}(b), we show the simulation and experimental results at three output power levels. We only vary the gain parameter to corresponding output power levels in all simulations shown in Fig. \ref{fig:duration}(a) and (b). These results also indicate that resonant dispersion does not contribute to the Tm:fiber amplifier dispersion, even with heavy doping concentration and strong gain per length. However, an interferometry measurement is still necessary as a conclusive proof. 

\newpage

\section{Carrier-envelope-offset frequency}
\begin{figure}[!ht]
    \renewcommand{\thefigure}{S\arabic{figure}} 
    \centering
    \includegraphics[width = 13cm]{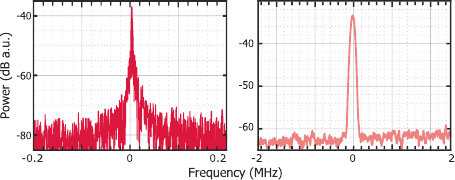}
	\caption{(a) f{\textsubscript {o}} at 1 kHz RBW (b) f{\textsubscript {o}} at 100 kHz RBW (right);  }
    \label{fig:fceo}
\end{figure}
With the 2-octave spectrum, retrieval of the carrier-envelope offset (CEO) frequency is straightforward. By focusing the laser output directly into a periodically-poled lithium niobate (PPLN), the second-harmonic-generation (SHG) provides the "2f" part. This SHG signal then beats with the dispersive wave at 1.05 µm giving the CEO frequency, f{\textsubscript{o}}. We used the same setup as our previous work \cite{Xing2020} and the PPLN is poled for doubling 2 µm light. Figure \ref{fig:fceo}(a) shows the f{\textsubscript{o}} at 1 kHz resolution bandwidth (RBW) with more than 40 dB signal-to-noise ratio. While our periodically-poled lithium niobate does not efficiently double 2.1 µm, we are still able to achieve 30 dB signal-to-noise [Fig. \ref{fig:fceo}(b)] at 100 kHz RBW.

\newpage

\section{More details on MIR frequency comb}
\begin{figure}[!ht]
    \renewcommand{\thefigure}{S\arabic{figure}} 
    \centering
    \includegraphics[width = 14cm]{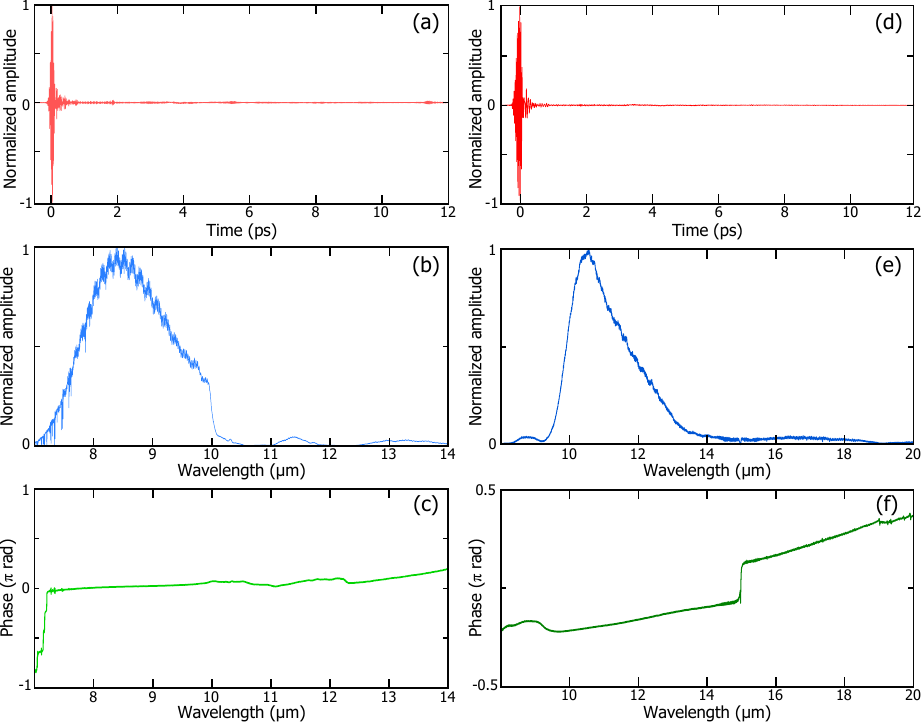}
	\caption{(a), (b) and (c) Field, spectrum and chirp of the MIR frequency comb from CSP crystal. Note the small replica at about 5.5 ps and 11 ps in (a) due to Fresnel reflection on the CSP surface. (d), (e) and (f) Field, spectrum and chirp of the MIR frequency comb from OP-GaAs crystal. The orientation pattern period is 79 µm.}
    \label{fig:EOS}
\end{figure}

Due to the Fresnel reflection at the CSP surface, a replica of the MIR electric field is expected. In addition, we expect to see interference pattern in the MIR spectrum. Figure \ref{fig:EOS}(a) shows the small replica signal at approximately 5.5 ps and 11 ps from the main electric field. The corresponding spectrum is plotted in linear scale in Fig. \ref{fig:EOS}(b) and clear interference can be observed. The sharp power drop at about 10 µm is due to the phonon absorption in CSP. Our OP-GaAs crystal has anti-reflection coating for 6$\sim$15 µm. For simplicity, we only discuss the OP-GaAs with 79 µm period here. Although the coating effectiveness starts to drop after 14 µm, the MIR power beyond 14 µm is very low and cannot lead to significant back-reflection. As a result, an electric field replica is absent in Fig. \ref{fig:EOS}(d) and Fig. \ref{fig:EOS}(e) shows no fringes. 

Since the majority of the MIR power does not overlap with atmosphere absorbent wavelength, we expect the pulse to be nearly transform-limited. Figure \ref{fig:EOS}(c) and (f) show the chirp of the MIR spectrum from CSP and OP-GaAs, respectively. Across the main part of both spectra there is minimal curvature of the spectral phase, indicating very little chirp. Chirp induced by water and CO\textsubscript{2} absorption lies at the edge of MIR spectrum for both crystals.

\nocite{*}
\bibliography{sl}%